\documentclass{PoS}
\usepackage{amsmath}
\title{Current status of dynamical modeling of fluctuations at the QCD phase transition in heavy-ion collisions}
\ShortTitle{Dynamical modeling}

\author{\speaker{Marlene Nahrgang}\\
        Department of Physics, Duke University, Durham, North Carolina 27708-0305, USA\\
        E-mail: \email{marlene.nahrgang@phy.duke.edu}}

\abstract{
 For a complete understanding of the QCD phase diagram it is important to connect first-principle thermodynamic calculations to experimental data from the RHIC Beam Energy Scan and the future experimental facilities FAIR, GSI, and NICA, Dubna. This can only be achieved by a realistic modeling of the dynamical evolution of critical fluctuations in heavy-ion collisions at the QCD phase transition. In this note I will summarize the current status of these dynamical models and highlight some of the important issues, which need to be addressed in the future.}
\FullConference{9th International Workshop on Critical Point and Onset of Deconfinement - CPOD2014,\\
		17-21 November 2014\\
		ZiF (Center of Interdisciplinary Research), University of Bielefeld, Germany}

\begin{document}
\section{Introduction}
Strongly interacting matter under extreme conditions such as high temperatures and high densities has been studied by the means of heavy-ion collision experiments for a couple of decades. With each new generation of accelerators the available beam energies increased steadily and with the restart of the LHC nuclear collisions at a center-of-mass energy per nucleon-nucleon pair $\sqrt{s_{\rm NN}}=5$~TeV will soon be possible. Due to the possibility of discovering a critical point in the phase diagram of QCD, however, a revived interest in lower beam energies is seen. By decreasing the beam energy one can vary the net-baryon density in the collision region due to the stopping of the incoming baryon currents of the two colliding nuclei. This idea is the central motivation of the Beam Energy Scan (BES) program at RHIC. Runs in 2010 and 2011 completed phase I by taking data at $\sqrt{s_{\rm NN}}=7.7,\, 11.5,\, 19.6,\, 27,\, 39,\, 62.4$ and $200$~GeV. In 2014 beam time for an additional run at $\sqrt{s_{\rm NN}}=14.5$~GeV was allocated. Phase II will run in 2018/2019 collecting data samples with much larger statistics.

First-principle lattice QCD calculations established that the phase transition at physical quark masses and zero baryochemical potential, $\mu_B=0$, is an analytic crossover \cite{Aoki:2006we,Borsanyi:2010bp}. At finite $\mu_B$ standard Monte Carlo importance sampling is not applicable anymore due to the fermionic sign problem. There is thus no exact method to solve QCD thermodynamics beyond $\mu_B=0$ that is currently feasible. Lattice QCD can be extended to values of $\mu_B/T$, which are not too large, by systematical Taylor-expansion of the pressure \cite{Borsanyi:2011sw,Bazavov:2012jq,Borsanyi:2012cr,Hegde:2014wga}. New approaches to solve QCD on the lattice at finite net-baryon density are being developed but have not yet reached the stage of quantitative statements about QCD thermodynamics \cite{Aarts:2014fsa}. 

Functional methods, like the Dyson-Schwinger equations (DSE) \cite{Fischer:2012vc,Fischer:2014ata} and the functional renormalization group (FRG) \cite{Pawlowski:2005xe,Pawlowski:2010ht}, can be applied in the nonperturbative regime of QCD and in the entire phase diagram without restrictions like the sign problem. The DSE are solved by using input from lattice QCD for the temperature-dependent quenched gluon propagator and a general ansatz for the quark-gluon vertex. This approach yields an excellent agreement with lattice QCD calculations in the crossover transition region at $\mu_B=0$. It predicts a critical point and a line of first-order phase transition at finite baryochemical potential  \cite{Fischer:2012vc,Fischer:2014ata}. In the FRG, the QCD flow equation for the thermodynamic potential is solved. The agreement with lattice QCD for the vacuum \cite{Braun:2014ata} and for pure Yang-Mills theory at finite temperatures is excellent \cite{Fister:2011uw}. The thermodynamics of low-energy effective models of QCD, see below, has also been solved by using FRG methods. For both of these functional methods the treatment of baryonic degrees of freedom needs to be improved in order to make reliable claims for values above $\mu_B/T\sim2$.

The qualitative structure of the crossover, critical point and first-order phase transition line at finite baryochemical potential has extensively been studied in chiral mesonic models, which include constituent quarks, such as the Quark-Meson (QM) or the Nambu-Jona-Lasino (NJL) model. Extending these models to include the Polyakov-loop improves the agreement with lattice QCD at $\mu_B=0$ and takes confinement into account by statistically suppressing one- and two-quark states. In these models the exact location of the critical point depends, however, on certain parameter choices and on the treatment of fluctuations, i.e. mean-field or beyond \cite{Schaefer:2007pw,Herbst:2010rf}.

In these proceedings I will make the case for the search of the critical point in a combined effort to connecting  first-principle calculations and experimental data via realistic modeling of the dynamical evolution of the matter created in heavy-ion collisions. These models must rely on input from first-principle calculations, include the relevant dynamics and nonequilibrium effects at the phase transition, and perform the analysis in similar manners as done for the experimental observables. The following sections are organized as follows:
In section \ref{sec:hiccpfo}, I outline the idea to discover the critical point and the first-order phase transition in heavy-ion collisions via proposed signatures. Section \ref{sec:dynmod} discusses the first attempts to describe the phase transition in a dynamical setup like a heavy-ion collision. Finally, in section \ref{sec:bes2}, I highlight some of the challenges for the BES phase II and summarize in section \ref{sec:summary}.

\section{Potential to discover the critical point and the first-order phase transition in heavy-ion collisions}\label{sec:hiccpfo}

In QCD, typical order parameters are the sigma field for the chiral phase transition, the Polyakov-loop for the confinement/deconfinement transition and the net-baryon density at finite baryochemical potential. It is expected that if the system created in a heavy-ion collision evolves along a trajectory through a critical point large event-by-event fluctuations in experimental observables like pion or proton multiplicities are seen~\cite{Stephanov:1998dy,Stephanov:1999zu}. This relies on the observation that in thermodynamically equilibrated systems, the correlation length of the fluctuations of the order parameter diverges, $\xi\to\infty$, at a second-order phase transition, which leads to the divergence of ensemble fluctuations. 
Of course in a spatially finite system, the correlation length can not exceed the size of the system. While this issue could be cured by finite-size scaling theory, it turns out that finite time plays a more crucial role in heavy-ion collisions. This is due to critical slowing down, when in addition to the correlation length also the relaxation time diverges and thus the closer the system gets to a critical point the longer it needs to achieve thermodynamic equilibrium. This becomes especially important for heavy-ion collisions which are highly dynamical~\cite{Berdnikov:1999ph}. Any fluctuation signal of the critical point will be weakened as a consequence. It was, thus, suggested to focus on higher-order moments of the measured particle distributions, which turned out to be more sensitive to the growth of the correlation length~\cite{Stephanov:2008qz}. 

The critical point can also be discovered indirectly by observing signals from a first-order phase transition. At a first-order phase transition the two phases have the same pressure and coexist due to the separation by the latent heat. When the collective expansion of the system created in a heavy-ion collision proceeds fast enough through a first-order phase transition it can be trapped in the meta-stable state below the actual transition temperature~\cite{Csernai:1995zn,Zabrodin:1998dk,Keranen:2002sw,Nahrgang:2011vn}. For small nucleation rates this supercooled state will decay via spinodal instabilities leading to the enhancement of low-momentum modes~\cite{Mishustin:1998eq,Randrup:2009gp,Randrup:2010ax}. 

One can see that for both scenarios, critical point and first-order phase transition, the effects of the dynamics are highly important. The nonequilibrium situation at the critical point weakens potential fluctuation signals, while nonequilibrium is necessary at the first-order phase transition in order to observe domain formation as a result of spinodal decomposition. For the understanding of the upcoming data from BES phase II, it will therefore be crucial to develop dynamical models, which include fluctuations at the phase transition.

\section{Dynamical models of the QCD phase transition}\label{sec:dynmod}
Modern dynamical models of heavy-ion collisions consist of the following three stages: the initial state of the incoming nuclei and the phase of preequilibrium evolution, the subsequent fluid dynamical evolution, which describes the expanding plasma phase and the phase transition, and the final hadronic cascade, which evolves the particles after particalization from the fluid via the Boltzmann transport equation until kinetic freeze-out \cite{Hirano:2005xf,Nonaka:2006yn,Petersen:2008dd,Werner:2012xh,Karpenko:2015xea}. Depending on the level of refinement these models are able to handle non-critical fluctuations, such as fluctuations in the initial state and final state. Including the phase transition into the fluid dynamical expansion is simple on the level of the equation of state and the transport coefficients. Including critical fluctuations at the QCD phase transition is, however, more challenging. In the following I will outline the current approaches to dynamical models of the phase transition.

\subsection{Nonequilibrium equation of state at the first-order phase transition}\label{sec:HQeosFOPT}
In \cite{Steinheimer:2012gc,Steinheimer:2013gla} a nonequilibrium equation of state was used in fluid dynamical simulations of heavy-ion collisions. This equation of state was constructed by joining a QGP equation of state via an explicit spline over the coexistence and spinodal region with a hadron gas equation of state. This is different from the usual procedure, where the two equations of state are joined by a Maxwell construction, which corresponds to the equilibrium situation, in which the coexistence and spinodal regions are ignored. The advantage of this approach is that the equation of state can easily be coupled to the deterministic fluid dynamical evolution and embedded in hybrid model simulations of heavy-ion collisions including the final hadronic cascade. Using the event-by-event initial conditions from UrQMD, it was seen that the initial inhomogeneities get amplified in the spinodal region of a first-order phase transition. Below the phase transition deep in the hadronic phase these amplified irregularities mostly decay before the kinetic freeze-out and the effect on observables is weakened. The quark-hadron phase transition features a coexistence region between dense quark matter and compressed nuclear matter, unlike the liquid-gas phase transition which has coexistence between dense quark matter and the vacuum \cite{Steinheimer:2013xxa}.

\subsection{Nonequilibrium chiral fluid dynamics}
The basic idea of nonequilibrium chiral fluid dynamics (N$\chi$FD) \cite{Nahrgang:2011mg,Nahrgang:2011mv,Nahrgang:2011vn,Herold:2013bi,Herold:2013qda,Herold:2014zoa} is to combine the explicit propagation of the order parameters with a fluid dynamical expansion of the bulk medium. The starting point is the QM or PQM model with constituent quarks. The quark degrees of freedom are integrated out via the two-particle irreducible action formalism and treated as a local heat bath for the propagation of the fluctuations of the order parameters, which is given by a Langevin equation (here for the sigma field)
 \begin{equation}
  \partial_\mu\partial^\mu\sigma+\frac{\delta U}{\delta\sigma}+g\rho_s+\eta\partial_t\sigma=\xi\, .
 \end{equation}
The following parameters are taken from the corresponding effective model: the vacuum potential $U$, the quark-meson coupling $g$ and the scalar density $\rho_s$.
 The damping coefficient $\eta$ is obtained from the interactions of the order parameter with the quarks and antiquarks and the variance of the noise field $\xi$ is given by the dissipation-fluctuation theorem as
\begin{equation}
\label{eq:dissfluctsigma}
 \langle\xi(t,\vec x)\xi(t',\vec x')\rangle=\delta(\vec x-\vec x')\delta(t-t')m_\sigma\eta\coth\left(\frac{m_\sigma}{2T}\right)\, ,
\end{equation}
where $m_\sigma$ is the local mass of the sigma field defined via the curvature of the potential.
 The heat bath is neither finite nor static but evolves according to the fluid dynamical equations. It is thus of importance to include the energy-momentum exchange between the fields of the order parameters and the fluid. This is achieved by including a source term in the fluid dynamical equation of the stress-energy tensor 
\begin{equation}
 \partial_\mu T^{\mu\nu}_{\rm q}= S^\nu=-\partial_\mu T^{\mu\nu}_\sigma\, ,\quad  \partial_\mu N^{\mu}_{\rm q}= 0\, .
\end{equation}
Due to the stochastic evolution of the order parameter, this source term makes the fluid dynamical fields stochastic quantities as well.

In N$\chi$FD fluctuations are generated dynamically which means that even for smooth initial conditions spinodal instabilities may arise during the evolution through the phase transition, as was shown in \cite{Herold:2013qda}. The equation of state as obtained from the (P)QM model is of the liquid-gas type, for which the pseudocritical pressure increases with temperature $\partial p_c/\partial T>0$ and vanishes at zero temperature. As a consequence dense quark matter coexists with the vacuum at zero temperature. In reality, however, dense quark matter as described at the end of the last section coexists with compressed nuclear matter. A proper treatment of the hadronic degrees of freedom at low temperatures is important for an inclusion of N$\chi$FD in a fully realistic dynamical model of heavy-ion collisions. 

Since the fluctuations in N$\chi$FD are dynamical and stochastic, unlike in the model described in section \ref{sec:HQeosFOPT}, where only initial irregularities are propagated deterministically, N$\chi$FD can also investigate effects at the critical point. It was shown that due to critical slowing down the correlation length of the fluctuations only grows up to $1.5-2.5$~fm \cite{Herold:2013bi} which is in good agreement with previous phenomenological studies \cite{Berdnikov:1999ph}. Although event-by-event fluctuations of the zero-mode of the sigma-field $\sigma_0$ can be studied in N$\chi$FD, we know that due to mixing with the net-baryon density $n_B$ at finite $\mu_B$ the full critical mode is a linear combination of $\sigma_0$ and $n_B$. Due to the diffusive nature of the dynamics of the net-baryon density, reflecting the conservation of baryon charge, $n_B$ becomes the true critical mode in terms of the long-time limit.

\subsection{Fluid dynamical fluctuations}
Conventional fluid dynamics propagates thermal averages, which are strictly defined only in infinite systems. In coarse-grained finite systems, the finite particle numbers lead to local thermal fluctuations around these averages. The general theory of fluid dynamic fluctuations \cite{landaulifschitz}  has recently been developed in the relativistic case for applications in heavy-ion collisions at top-RHIC and LHC energies \cite{Kapusta:2011gt,Murase:2013tma,Young:2014pka}. 
In the context of heavy-ion collisions it has been realized that the data on flow observables is better described by including small viscous corrections to the ideal fluid dynamical simulations. In linear response theory the viscosities can be obtained from retarded correlators of the stress-energy tensor via the dissipation-fluctuation theorem. These are the so-called Kubo-relations. It also tells us that the fast processes that lead to local equilibration in a fluid lead to noise. As a consequence, fluctuations need to be directly included in the fluid dynamical fields and propagated by the fluid dynamical equations.

In stochastic viscous fluid dynamics the stress-energy tensor and the net-baryon current have an equilibrium, a viscous and a stochastic component
\begin{align}
 T^{\mu\nu}&=T^{\mu\nu}_{\rm eq}+\Delta T^{\mu\nu}_{\rm visc}+\Xi^{\mu\nu}\, ,\\
 N^{\mu}&=N^{\mu}_{\rm eq}+\Delta N^{\mu}_{\rm visc}+{\rm I}^{\mu}\, .
\end{align}
Since $\langle\Xi^{\mu\nu}\rangle=0$ and $\langle{\rm I}^{\mu}\rangle=0$, single-particle quantities are not changed and the two formulations, conventional versus stochastic viscous fluid dynamics, differ only when one calculates correlation functions. The correlation functions of the noise contributions can be derived in linearized fluid dynamics and yield
\begin{align}
 \langle\Xi^{\mu\nu}(x)\Xi^{\alpha\beta}(x')\rangle&=2T\left(\eta\left(\Delta^{\mu\alpha}\Delta^{\nu\beta}+\Delta^{\mu\beta}\Delta^{\nu\alpha}\right)+\left(\zeta-\frac{2}{3}\eta\right)\Delta^{\mu\nu}\Delta^{\alpha\beta}\right)\delta^{(4)}(x-x')\, ,\\
 \langle{\rm I}^{\mu}(x){\rm I}^{\alpha}(x')\rangle&=2T\kappa\,\Delta^{\mu\alpha}\delta^{(4)}(x-x')\, ,
\end{align}
with the shear and bulk viscosities, $\eta$ and $\zeta$, and the baryon charge conductivity, $\kappa$, and the usual fluid dynamical projector $\Delta^{\mu\nu}$. The $\delta^{(4)}(x-x')$ functions indicate that these expression are in the white noise approximation. 

\section{Challenges for the Beam Energy Scan Phase II}\label{sec:bes2}
In addition to advancing the modeling of the dynamical fluctuations at the QCD phase transition there remain other challenges for both theory and experiment in order to understand the existing and upcoming data from heavy-ion collision at finite baryon density by linking QCD thermodynamics to experiment. I will now discuss the most important aspects:
\begin{itemize}
 \item[\textbullet] \textit{Equation of state and transport coefficients}: The equation of state and transport coefficients are necessary input for the modeling of heavy-ion collisions. By including dynamical fluctuations at the phase transition one tries to achieve consistency in thermodynamic quantities, $p=p(e,n_B)$, susceptibilities, $\chi_n=\partial^n (p/T^4)/\partial (\mu_B/T)^n$, and the transport coefficients, $\eta$, $\zeta$ and $\kappa$, as functions of temperature and baryochemical potential. As explained above for the equation of state (quark-hadron versus liquid-gas phase transition) the different choices can have significant phenomenological consequences. At small baryochemical potential the most reliable input is expected from lattice QCD calculations for the thermodynamic quantities. At finite baryochemical potential and for the dynamical transport properties approaches like DSE and FRG need to be developed further. Chiral effective models including the quark-hadron phase transition \cite{Hempel:2013tfa} can in the meantime be investigated.
 \item[\textbullet] \textit{Initial state and final state}: Fluctuations in the initial state and the final state fluctuations are sources of fluctuations that do not relate to the QCD phase transition and are important to be understood as a benchmark quantity. Here, the fluctuations in the initial baryon transport and stopping are of particular interest but only insufficiently studied up to now. Additional experimental data of correlations and fluctuations in rapidity will help understanding this contribution.
 Theoretically, at the transitions from the initial state to fluid dynamical fields and from these fields to final particles via particalization correlations and fluctuations should be preserved. At the current level, initial thermalization is assumed ad hoc rather than demonstrated dynamically and quantities are mapped on averaged fluid dynamical fields thus throwing away correlations, which might have built up initially. Similarly, at particalization particle production occurs according to single-particle distribution functions, which again do not preserve the correlations and fluctuations which developed during the previous stage. For a proper understanding of the contributions from all stages consistent transition models need to be built. 
 \item[\textbullet] \textit{Net-baryon versus net-proton fluctuations}: 
Considering that net-baryon number density is the true critical mode which imprints signals for the critical point directly in the net-baryon number, one needs to be aware that experimentally only protons and antiprotons are measured. Due to isospin randomizing processes, e.g. final state nucleon-pion scatterings via a $\Delta$-resonance, critical fluctuations are significantly obscured in the net-proton number fluctuations compared to net-baryon number. If complete isospin randomization is achieved, i.e. for high pion densities and a sufficiently long final hadronic stage, net-baryon number fluctuations can be reconstructed from measured proton- and antiproton distributions as derived in~\cite{Kitazawa:2011wh,Kitazawa:2012at}. Benchmark calculations in hadron resonance gas models can include these effects statistically~\cite{Nahrgang:2014fza}. It remains an open question down to which beam energies full isospin randomization is indeed achieved. The missing contribution from strange baryons might as well add a dependence on $\sqrt{s}$ and the quantitative effect of it needs to be determined carefully.
  \item[\textbullet] \textit{Global charge conservation}: In heavy-ion collisions event-by-event fluctuations in the net-baryon number, net-electric charge or net-strangeness number are observed only because the detectors do not measure all particles in the full phase space. With varying beam energy due to baryon stopping the average number of net-baryons seen in a certain rapidity window changes. It was shown in transport model studies, which describe the microscopic nature of individual scatterings, that while net-baryon number fluctuations change significantly toward lower $\sqrt{s}$ due to reaching the limit of global charge conservation, net-proton number remains mostly unaffected~\cite{Schuster:2009jv}. Any experimental cut affects fluctuation observables, since observing only a fraction of all net-charges biases the distribution toward Poisson-like behavior even if the underlying distribution is far from Poisson~\cite{Bzdak:2012an}.
\item[\textbullet] \textit{Efficiency corrections}: Since the detectors only have a limited reconstruction efficiency additional fluctuations will affect all moments of the measured distributions~\cite{Bzdak:2012ab}. In principle, these reconstruction formulae can be applied, which might even be momentum dependent~\cite{Bzdak:2013pha}. This will be a particularly important issue for the extended transverse momentum coverage after the STAR collaboration uses two detectors, the Time Projection Chamber (TPC) and the Time-of-Flight (ToF) for particle identification. Experimental errors are also affected by efficiency corrections~\cite{Luo:2014rea}.
 \item[\textbullet] \textit{Phase transition and freeze-out conditions}: Nonmonotonic fluctuation pattern as a function of $\sqrt{s}$ could also be generated by an interplay of dynamical fluctuations from the phase transition and the location of freeze-out conditions \cite{Alba:2014eba}. It will be necessary to apply models in the whole QCD phase diagram in order to distinguish between fluctuations in the crossover regime~\cite{Redlich:2012xf}, the critical region or the first-order phase transition. Even higher-order cumulants might play an important role here.
\end{itemize}

\section{Summary}\label{sec:summary}
In these proceedings I have outlined the importance of the realistic modeling of the dynamical fluctuations at the QCD phase transition in heavy-ion collisions. Without this bridge between first-principle calculations of QCD thermodynamics and the experimental data of previous, current and upcoming heavy-ion collisions a full understanding of the QCD phase diagram will not be achieved.

Currently, the most reliable models of bulk observables in heavy-ion collisions exist for high beam energies, where thermodynamical input in form of the equation of state can directly be obtained from lattice QCD calculations. Toward lower beam energies and higher baryonic densities the uncertainties from the initial state, the equation of state and the transport coefficients become larger. On top of the existing models, a fully dynamical description of the fluctuations at the QCD phase transition needs to be developed. In section \ref{sec:dynmod}, I presented current attempts to include the dynamics of critical fluctuations and instabilities at the QCD critical point and the first-order phase transition. They go into the right direction and reproduce already important features like critical slowing down and spinodal decomposition, but are not yet flexible and complete enough to be coupled consistently to bulk models of the underlying noncritical dynamics.

Concerted effort will be necessary ranging from first-principle QCD thermodynamics, to the development of bulk models including noncritical sources of fluctuations, to the work of experimental collaborations and to the dynamical modeling of fluctuations at the QCD phase transition.

 \section*{Acknowledgments}
The author acknowledges support from the U.S. Department of Energy under grant DE-FG02-05ER41367 and a fellowship within the Postdoc-Program of the German Academic Exchange Service (DAAD).

\end{document}